# Glassy Nature of $Ce_{65}Al_{25}Co_{10}$ Alloy: A Metallic Glass

Mohd Alam, Dharmendra Singh, O.N. Srivastava, R. S. Tiwari

Department of Physics, BHU, Varanasi - 221005

## Abstract

In the present investigation the glass forming ability of $Ce_{65}Al_{25}Co_{10}$ metallic glass has been reported. It was successfully synthesised using melt spinning techniques. The existence of amorphous phase in $Ce_{65}Al_{25}Co_{10}$ alloy have been proved using X-ray diffraction (XRD) and transmission electron microscopy (TEM) measurements. The thermal analysis of the synthesised sample using differential scanning calorimetry (DSC) shows a glass transition around 371 K, thus confirms existence of glassy phase at room temperature. The hardness of the synthesised sample at different load have been demonstrated. Moreover, the yield strength of the synthesised sample has also been calculated by means of hardness data and Meyer exponent.

## 1. Introduction

The increasing demand of glassy materials for technological application drives people's attentions towards metallic glass (MG), which has many exclusive properties for the materials research [1, 2]. The earlier development of glass formation in binary Cu-Zr alloys has attracted towards somewhat simple and easy glass-forming materials, like invention of the of scarce glass-forming capability in ternary Cu- Zr-Al as well as in quaternary Cu-Zr-Al-Y alloys [1, 3-5]. The ductileness of bulk-metallic-glasses (BMGs) around room temperature was poorly low though it's unveiled the near theoretical strength before failure showing partial/absence of dislocational plasticity due to highly localization of the deformation movements [6, 7]. The example of localized deformation is the shear banding (which is a arrangement of extreme localization of the plastic deformation in the system) due to the accumulation of plastic strain within a narrow region revealing strain/thermal softening [8]. To confine the propagation of shear bands and increasing ductility of MGs, heterogenous structural has been developed whose example can be found in the report of J. Das et. al., [1].

Furthermore, numerous rare-earth-based MG has been reported in literature [9-11]. Cerium based alloys has prompted a lot of interest owing to their unusual glass-forming capability with low glass transition temperature, super-plasticity, interesting magnetic & and mechanical properties, thermo-plasticity near room temperature in super-cooled liquid region and leads a potential candidate for technological applications as a glassy material [9-15]. Based on kinks in amorphous XRD (X-xay diffraction) pattern, a polymorphic transition has been observed in MG $Ce_{55}Al_{45}$ and in BMG $La_{32}Ce_{32}Al_{16}Ni_5Cu_{15}$ due to the delocalization of Ce-4f electrons under high pressure [16-18]. Similar electronic polymorphisms, attributed to Ce constituent, in $Ce_{75}Al_{25}$ metallic glass was also proved using pressure dependent XAS (x-ray absorption spectroscopy) [2].

The glassy or amorphous alloys has been formed via different methods; however much efficient method is quenching with a high rate (rapid solidification) from a liquid phase using different techniques viz melt spinning, liquid forging, Cu-mold casting etc [19]. Since crystal nucleation needs a sufficiently low cooling rate, thus rapid cooling leads the system to falls out from equilibrium and the temperature at which this happens known as $T_g$ (glass transition temperature) [20]. Thus, the MGs passes a solid physique but their behaviour is liquid-like. It has been highlighted that the glass-forming-ability (GFA) is strongly influenced by the packing density (PD) of non-crystalline martials and the importance of the atomic PD in the formation of MGs was discussed recently [21, 22]. However, successive three-dimensional massive glassy alloys i.e., BMGs also has been produced using several casting techniques of the size of greater than 1 mm up to 72 mm [23-26].

In this report, we are going to present a brief description of MGs by characterizing $Ce_{65}Al_{25}Co_{10}$. We have carried out XRD and TEM measurements to describe possible glassy phase. Further, to see glass transition temperature we have carried out DSC measurement. Additionally, the mechanical behaviour has been studies using indentation micrographs of the $Ce_{65}Al_{25}Co_{10}$ MGs and have been reported here.

## 2. Experimental details

The MG $Ce_{65}Al_{25}Co_{10}$ was synthesised using melt spinning techniques as described in earlier report. The highly pure elements of constituent elements were taken in stoichiometric ratio. They were mixed and melted in a silica crucible via induction furnace. After cooling we got ingot of mixture and the ingot was remelted for several times (5 to 10 times) to make sure completely homogeneous mixture. A part of this ingot was placed in a silica nozzle with a circular hole at the bottom of around 1 mm diameter. Again, this was melted using induction and melt of ingots falls on a rotating copper wheel (around 14 cm diameter) with a speed of around 40 m/s and thus ingots were converted into ribbons with a length of few cm (1 to 4) and that of length of around 40 μm. The melting and ribbons formation was done under inert atmosphere (under highly pure Argon environment at 60 kPa pressure) to avoid oxidation. The XRD spectrum was collected using X'Pert Pro diffractometer, where Cu-target was used for x-ray generation working at a potential of 25 kV with a tube current of 20 mA. For investigation under transmission electron microscopy (using Technai 20$G^2$ electron microscope operating at a potential of 200 kV), a part of as-synthesized ribbons was thinned with the help of an electrolyte (a solution of 70 percent methanol and 30 percent nitric acid) at -20 ºC. The

SHIMADZU DSC-60 was used for the thermal analysis of the samples with a heating rate of 20 K/min. The HMV-2 T microhardness tester was used for microhardness measurements of the sample at different load.

## 3. Results and discussions

**3.1 Structure analysis:**

Fig. 1 shows XRD patterns of $Ce_{65}Al_{25}Co_{10}$ melt spun ribbons. The XRD pattern does not shows any characteristic sharp peaks, suggesting absence of long rang ordering in the sample. A broad single broad maximum observed in XRD pattern (which is only due to some short-range ordering), within the angular range 28º–37º, is a characteristic of glassy/amorphous materials [11]. Fig. 2 (a) shows the transmission electron microscope (TEM) micrograph displaying a contrast free region and Fig. 2 (b) presents the corresponding SAD (selected area diffraction) pattern of $Ce_{65}Al_{25}Co_{10}$ alloy, showing diffuse halos. This is the characteristics of the absence of residual contrast in the bright field image and thus confirms the presence of a homogenous glassy/amourphous phase. Generally, any approach/description for the amorphous structure tells that it is a homogeneous isotropic structure, factually it cannot be always true. In the cases where these alloys contains more than one metals with similar scattering amplitude, the emergence of of inhomogeneity/different amorphous phases appears as different chemical species cause appearance of at least two types of shortest-range ordering among atoms, thus unsurprisingly results in phase separation [27, 28]. The phase separation in MGs has been investigated by many authors earlier [29]. However, for the MGs having negative heat of mixing, the phase separation can be understood with the help of a model based on nature of geometrical clusters [30]. This model supposed that the MGs have long-range translational ordering with geometrical mismatch in main clusters which are connected through intermediate clusters known as cementing/glue cluster [31-34].

### 3.2 Thermal analysis:

Fig. 3 consists of a typical DSC profile of $Ce_{65}Al_{25}Co_{10}$ MG, which is generally used for the thermal analysis of as-synthesised samples. A smeared endothermic peak can be seen around 371 K, which is eventually the glass transition temperatures ($T_g$). Further, onset crystallization temperature ($T_x$), peak crystallization temperature ($T_p$) and supercooled liquid region ($\Delta T_x$) have been summarized in table 1. The larger supercooled liquid region $\Delta T_x$, means more stable supercooled liquid against crystallization and thus better GFA [35]. These are the characteristics temperatures used for characterization of GFA of MGs, and based on these temperatures including melting temperature ($T_m$), different GFA parameters have been in recent years [36]. The most common parameters have been reported as the reduced glass transition temperature, $T_{rg}$ ($T_g/T_m$) [37] and supercooled liquid region $\Delta T_x$ ($T_x - T_g$) [35]. The MGs with the value of $T_{rg}$ within 0.66–0.69, have been verified to have better GFA [36]. For example, based on above criteria, it is reported that the Ce–Al–Cu based alloys has better GFA than that of the Ce–Al–Co [36]. Lu and Liu proposed a new GFA parameter for BMGs depending on crystallization developments, which is $\gamma = T_x/(T_g + T_m)$ and the value of $\gamma$ lies in the range 0.35 to 0.50 exhibits a good relationship with the GFA for the most of the MGs and some oxide glasses as well. However, classical nucleation and growth theory established other parameters to evaluate GFA of an alloy, which is $\delta = T_x/(T_m - T_g)$ [38] and $\beta = T_xT_g/(T_m + T_x)^2$ [39]. Further, a thermodynamic parameter for GFA, $\omega = T_m(T_m + T_x)/(T_x(T_m - T_x))$, has been proposed by Ji and Pan, by means of Gibbs free energy difference between crystal and liquid. Ji and Pan believed that the $\omega$ is the most reliable parameter applicable for GFA among above-described parameters [40].

Earlier, the molecular dynamics (MD) simulations have been performed by Lü, and W. H. Wang to see the glass transition temperature of CuZr alloy at zero pressure [41]. They successively reported the single-particle dynamics of the alloy, around the glass-transition temperature, $T_g$. Cooling to a temperature lower than 1.2 $T_g$, the molecular dynamics becomes heterogeneous, which was proved by the cage-jump motion and the liquid falls out of equilibrium when it was cooled below 1.02 $T_g$. Using the thermal variation of enthalpy calculation in the simulations, the $T_g$ was found around 720 $\pm$ 5K, however it was higher than the experimental value of around 670 K [41, 42].

### 3.3 Indentation characteristics:

Fig. 4. shows the demonstrative optical micrographs of indentation marks for the as-synthesized ribbons of $Ce_{65}Al_{25}Co_{10}$ alloy. The clear wave-like shear bands (marked by black arrows) around the indentation edge have been observed, which is a characteristic of the MGs [11]. These bands are basically a plastic-deformation arising in the system due to indentation. In MGs, the shear-bands are of great important as they control the plasticity and failure of the materials. It has been reported by Tang et al., that there is rise in temperature in the shear band of a simulated MG, which was evidenced by MD simulations [43]. Experimentally, it is difficult to measure temperature rise ($\Delta T$) in shear band due to spatial and temporal localization [43-45]. However, using MD simulations, a spectrum of $\Delta T$ was recorded and it was confirmed that the $\Delta T$ depends on both strain rate and sample size of MGs under tension and the $\Delta T$ is correlated with maximum sliding velocity of the shear bands [43].

As evidenced from the Fig. 4 (a-c), a regular and crack free indentation impressions occurs in $Ce_{65}Al_{25}Co_{10}$ MG, at the load up to 100 g. However, it was cracked was at 500 g. The hardness (VHN) was calculated (Table 2) using the formula [46];

$$VHN = 1.854 * 9.8 * \frac{P}{d^2} \quad \text{GPa} \quad (1)$$

where, P and d is the is the load in g and diagonal length in μm respectively. Interestingly, the hardness value of $Ce_{65}Al_{25}Co_{10}$ alloy is larger than that of constituent's metal (as standard value of hardness for Ce is 0.21 to 0.47 GPa, for Al is 1.17 GPa and for Co it is 0.143 GPa) and also it is larger than the $Ce_{60}Cu_{25}Al_{15}$ (~2.54 GPa). Fig. 5 shows the variation of hardness with load. It was observed that the hardness decreases at higher load, similar to other reported MGs, because of indentation size effect due to flow of the materials under load. By means of hardness value we can calculate the 0.2% offset yield strength. The method of the calculation of offset yield strength has been described in mentioned in many report [47]. For this purpose, we need to determine Meyer exponent (n), given by the equation

$$P = Kd^n \qquad (2)$$

Where P is the load and d is the diagonal length (Table 3), as described earlier. The value of n can be estimated from the slope of log P versus log d plot and the intercept gives the value of K (which is a constant for the material related to the resistance against indenter penetration. Fig. 6 shows the log P vs. log d plot of $Ce_{65}Al_{25}Co_{10}$ alloy. The value of the yield strength is given by

$$\sigma_0 = \left[\frac{VHN}{3}\right](0.1)^n \qquad (3)$$

Which gives the value of 0.2% offset yield strength around 1.21 GPa $Ce_{65}Al_{25}Co_{10}$ MG at 100 g load.

## 4. Conclusion

In summary, a cerium based metallic glass $Ce_{65}Al_{25}Co_{10}$ metallic glass was prepared using melt spinning technique via rapid cooling. The XRD analysis shows a single broad diffraction maximum within the angular range 28°–37°, without any detectable sharp peaks which is eventually a characteristic of amorphous phase. TEM micrograph shows clear contrast free (the absence of residual contrast in the bright field image) region and corresponding selected area diffraction pattern (SADP) demonstrates existence of diffuse halos, which again supports amorphous/glassy phase, because of the absence of periodic arrangements of the atoms or lack of long-range ordering. the DSC spectrum shows a smeared endothermic peak of the glass transition around 371 K. The hardness measurement using indentation shows that the hardness decreases with increase in the load due to indentation size effect. The Vickers's hardness (VHN) and yield strength for the $Ce_{65}Al_{25}Co_{10}$ alloy was been found interestingly large which is around 2.81 and 1.21 GPa respectively. The hardness value of $Ce_{65}Al_{25}Co_{10}$ alloy is found to larger than the constituent's metal. Optical micrographs of indentation mark display the wave- like patterns (similar to waves in liquid when we drop of a piece of stone) around the indentation periphery disclose the establishment of shear bands in sample. From these micrographs it can be seen that the indentation impressions are consistent and crack free at a load up to 100 g and a significant crack was observed at 200 g.

References:

1. Jayanta Das, Mei Bo Tang, Ki Buem Kim, Ralf Theissmann, Falko Baier, Wei Hua Wang, and Jurgen Eckert, "Work-Hardenable" Ductile Bulk Metallic Glass, PRL 94, 205501 (2005).
2. Qiao-shi Zeng, Yang Ding, Wendy L. Mao, Wenge Yang, Stas.V. Sinogeikin, Jinfu Shu, Ho-kwang Mao, and J. Z. Jiang, Origin of Pressure-Induced Polyamorphism in Ce75Al25 Metallic Glass, PRL 104, 105702 (2010).
3. D. Wang, Y. Li, B. B. Sun, M. L. Sui, K. Lu, and E. Ma, Appl. Phys. Lett. 84, 4029 (2004).
4. A. Inoue and W. Wang, Mater. Trans., JIM 43, 2921 (2002).
5. D. Xu, G. Duan, and W. L. Johnson, Phys. Rev. Lett. 92, 245504 (2004).
6. F. Spapen, Acta Metall. 25, 42 (1977)

7. A. L. Greer, Science 267, 1947 (1995).

8. H. Chen, Y. He, G. J. Shiftlet, and S. J. Poon, Nature (London) 367, 541 (1994).

9. B. Zhang, R.J. Wang, D.Q. Zhao, M.X. Pan, W.H. Wang, Properties of Ce-based bulk metallic glass-forming alloys, Phys. Rev. B 70 (2004) 224208.

10. Q.S. Zeng, C.R. Rotundu, W.L. Mao, J.H. Dai, Y.M. Xiao, P. Chow, X.J. Chen, C.L. Qin, H.K. Mao, J.Z. Jiang, Low temperature transport properties of Ce–Al metallic glasses, J. Appl. Phys. 109 (2011) 113716.

11. Dharmendra Singh, R.K. Mandal, O.N. Srivastava, R.S. Tiwari, Glass forming ability, thermal stability and indentation characteristics of $Ce_{60}Cu_{25}Al_{15} xGa_x$ ($0 \leq x \leq 4$) metallic glasses, Journal of Non-Crystalline Solids 427 (2015) 98–103.

12. D.F. Franceschini, S.F. da Cunha, Magnetic properties of $Ce(Fe_{1-x}Al_x)_2$ for $x \leq 0.20$, J. Magn. Magn. Mater. **51** (1985) 280–290.

13. D.H. Xiao, J.N. Wang, D.Y. Ding, H.L. Yang, Effect of rare earth Ce addition on the microstructure and mechanical properties of an Al–Cu–Mg–Ag alloy, J. Alloys Compd. 352 (2003) 84–88.

14. B. Zhang, D.Q. Zhao, M.X. Pan, R.J. Wang, W.H. Wang, Formation of cerium-based bulk metallic glasses, Acta Mater. **54** (2006) 3025–3032.

15. Q. Luo, D.Q. Zhao, M.X. Pan, W.H. Wang, Magnetocaloric effect in Gd-based bulk metallic glasses, Appl. Phys. Lett. **89** (2006) (081914-1-3).

16. H.W. Sheng et al., Polyamorphism in a metallic glass, Nature Mater. 6, 192 (2007).

17. A. R. Yavari, The changing faces of disorder, Nature Mater. 6, 181 (2007).

18. Q. S. Zeng et al., Anomalous compression behavior in lanthanum/cerium-based metallic glass under high pressure, Proc. Natl. Acad. Sci. U.S.A. 104, 13565 (2007).

19. Dmitri V. Louzguine-Luzgin, Ichiro Seki, Tokujiro Yamamoto, Hitoshi Kawaji, C. Suryanarayana, and Akihisa Inoue, Double-stage glass transition in a metallic glass, PHYSICAL REVIEW B **81**, 144202 (2010).

20. L. Berthier and M. D. Ediger, Phys. Today 69(1), 40 (2016).

21. D. B. Miracle, W. S. Sanders, and O. N. Senkov, Philos. Mag. 83, 2409 (2003).

22. H. W. Sheng, W. K. Luo, F. M. Alamgir, J. M. Bai, and E. Ma, Nature (London) 439, 419 (2006).

23. A. Inoue, Mater. Trans., JIM 36, 866 (1995).

24. W. L. Johnson, MRS Bull. 24, 42 (1999).

25. A. Inoue, Acta Mater. 48, 279 (2000).

26. C. Suryanarayana and A. Inoue, Bulk Metallic Glasses (CRC Press, Boca Raton, FL, 2010.

27. Kim CO, Johnson WL. Amorphous phase separation in the metallic glasses (Pb1-ySby)1−xAux. Physical Review B.23,143 (1991).

28. Cao Q , Li J, Zhou Y, Jiang J. Mechanically driven phase separation and corresponding microhardness change in Cu60Zr20Ti20 bulk metallic glass. Applied Physics Letters 86, 081913 (2005).

29. Park BJ, Chang HJ, Kim DH, Kim WT. In situ formation of two amorphous phases by liquid phase separation in Y-Ti-Al-Co alloy. Applied Physics Letters. 85(26):6353 (2004).

30. Miracle DB. The efficient cluster packing model – An atomic structural model for metallic glasses. Acta Materialia. 2006; 54: 4317-4336.

31. Singh D, Mandal RK, Tiwari RS, Srivastava ON. Nanoindentation characteristics of Zr69.5Al7.5−xGaxCu12Ni11 glasses and their nanocomposites. Journal of Alloys and Compounds. 2011;509:8657-8663.

32. Singh D, Basu S, Mandal RK, Srivastava ON, Tiwari RS. Formation of nano-amorphous domains in Ce75Al25 − xGax alloys with delocalization of cerium 4f electrons. Intermetallics.2015;67:87-93.

33. Singh D, Singh D, Mandal RK, Srivastava ON, Tiwari RS. Glass forming ability, thermal stability and indentation characteristics of Ce75Al25 − xGax metallic glasses. Journal of Alloys and Compounds. 2014;590:15-20.

34. Sohn SW, Yook W, Kim WT, Kim DH. Phase separation in bulk type Gd-Zr-Al-Ni metallic glass. Intermetallics. 2012;23:57-62.

35. A. Inoue, T. Zhang, T. Masumoto, Zr–Al–Ni Amorphous Alloys with High Glass Transition Temperature and Significant Supercooled Liquid Region, Mater. Trans. JIM 31 (1990) 177–183.

36. C. Tang, W. Pan, B. Zhang, J.Wang, H. Zhou, Investigation of glass forming ability in Ce–Al–Co (Cu) alloys, J. Non-Cryst. Solids 383 (2014) 6–12.

37. D. Turnbull, Under what conditions can a glass be formed? Contemp. Phys. 10 (5) (1969) 473–488.

38. Q.J. Chen, J. Shen, D. Zhang, H.B. Fan, J.F. Sun, D.G. McCartney, A new criterion for evaluating the glass-forming ability of bulk metallic glasses, Mater. Sci. Eng., A 433 (2006) 155–160.

39. Z.Z. Yuan, S.L. Bao, Y. Lu, D.P. Zhang, L. Yao, A new criterion for evaluating the glass-forming ability of bulk glass forming alloys, J. Alloys Compd. 459 (2008) 251–260.

40. X.L. Ji, Y. Pan, A thermodynamic approach to assess glass-forming ability of bulk metallic glasses, Trans. Nonferrous Met. Soc. China 19 (5) (2009) 1271–1279.

41. Y. J. Lü, and W. H. Wang, Single-particle dynamics near the glass transition of a metallic glass, PHYSICAL REVIEW E 94, 062611 (2016).
42. M. B. Tang, D. Q. Zhao, M. X. Pan, and W. H. Wang, Binary Cu–Zr Bulk Metallic Glasses, Chin. Phys. Lett. 21, 901 (2004).
43. Chunguang Tang, Jiaojiao Yi, Wanqiang Xu, and Michael Ferry, Temperature rise in shear bands in a simulated metallic glass, PHYSICAL REVIEW B 98, 224203 (2018).
44. A. J. Cao, Y. Q. Cheng, and E.Ma, Structural processes that initiate shear localization in metallic glass, Acta Mater. 57, 5146 (2009).
45. N. P. Bailey, J. Schiotz, and K. W. Jacobsen, Atomistic simulation study of the shear-band deformation mechanism in Mg-Cu metallic glasses, Phys. Rev. B 73, 064108 (2006).
46. N.K. Mukhopadhyay, G.C. Weatherly, J.D. Embury, An analysis of microhardness of single-quasicrystals in the Al–Cu–Co–Si system, Mater. Sci. Eng. A 315 (2001) 202–210.
47. J.R. Cahoon, W.H. Broughton, A.R. Kutzak, The determination of yield strength from hardness measurements, Metall. Trans. 2 (1971) 1979–1983.

**Table 1:** Thermal analysis of the melt-spun $Ce_{65}Cu_{10}Al_{25}$ ribbon, where $T_g$: glass transition temperature; $T_x$: crystallization temperature; $T_p$: exothermic peak; $\Delta T_x$: supercooled liquid region.

| $T_g$ (K) | $T_x$ (K) | $T_p$ (K) | $\Delta T_x = T_x - T_g$ |
|---|---|---|---|
| 378 | 481 | 501 | 103 |

**Table 2:** The values of VHN number, diagonal length at different loads and corresponding VHN in GPa.

| Load (in gram) | VHN number | d (Average diagonal length in μm) | VHN (in GPa) |
|---|---|---|---|
| 25 | 311 | 12.22 | 3.05 |
| 50 | 303 | 17.50 | 2.97 |
| 100 | 287 | 25.41 | 2.81 |

**Table 3:** Log tables of load (P) and their diagonal lengths (d).

| P (load in Kg) | d (in mm) | Log P (y-axis) | Log d(x-axis) |
|---|---|---|---|
| 0.025 | 0.01222 | -1.6020 | -1.91290 |
| 0.050 | 0.01750 | -1.3010 | -1.75696 |
| 0.100 | 0.02541 | -1.0000 | -1.59499 |

*Figure captions:*

**Figure 1:** XRD patterns of as synthesized ribbons of $Ce_{65}Al_{25}Co_{10}$ alloy.

**Figure 2:** (a) A representative bright field TEM microstructure of $Ce_{65}Al_{25}Co_{10}$ alloy alloys for (b) The corresponding selected area diffraction pattern.

**Figure 3:** DSC spectrum of as synthesized ribbons of $Ce_{65}Al_{25}Co_{10}$ alloy.

**Figure 4:** Nature of indentation for the as-synthesized ribbons of $Ce_{65}Al_{25}Co_{10}$ alloy at (a) 25 g (b) 50 g and (c) 100 g loads.

**Figure 5:** Variation of hardness (VHN) with respect to load (g) for the as synthesized ribbons of $Ce_{65}Al_{25}Co_{10}$ alloy.

**Figure 6:** Log P vs. Log d plots for the as synthesized ribbons of $Ce_{65}Al_{25}Co_{10}$ alloys alloy.

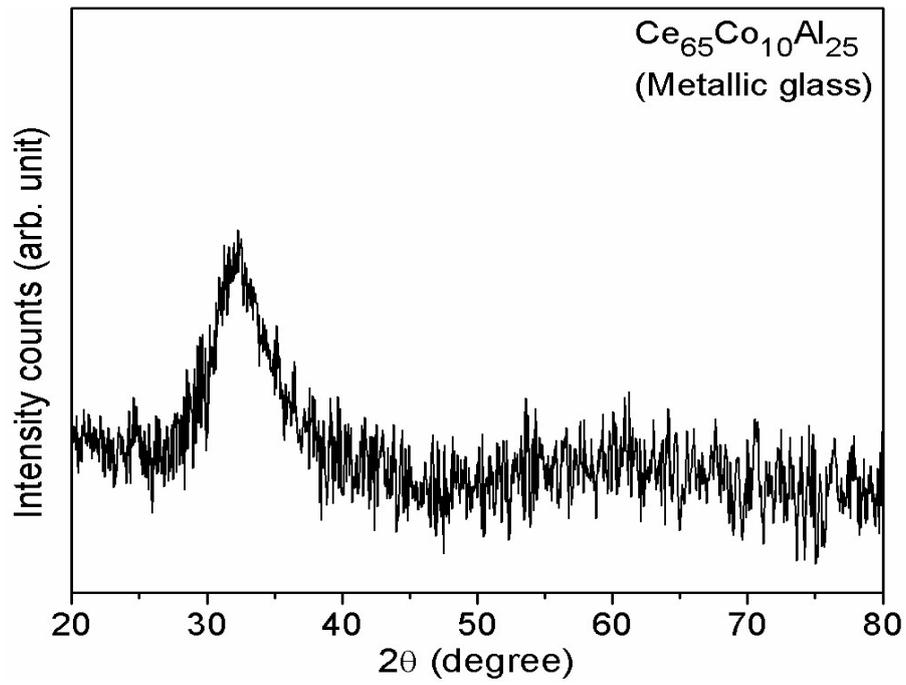

Figure 1

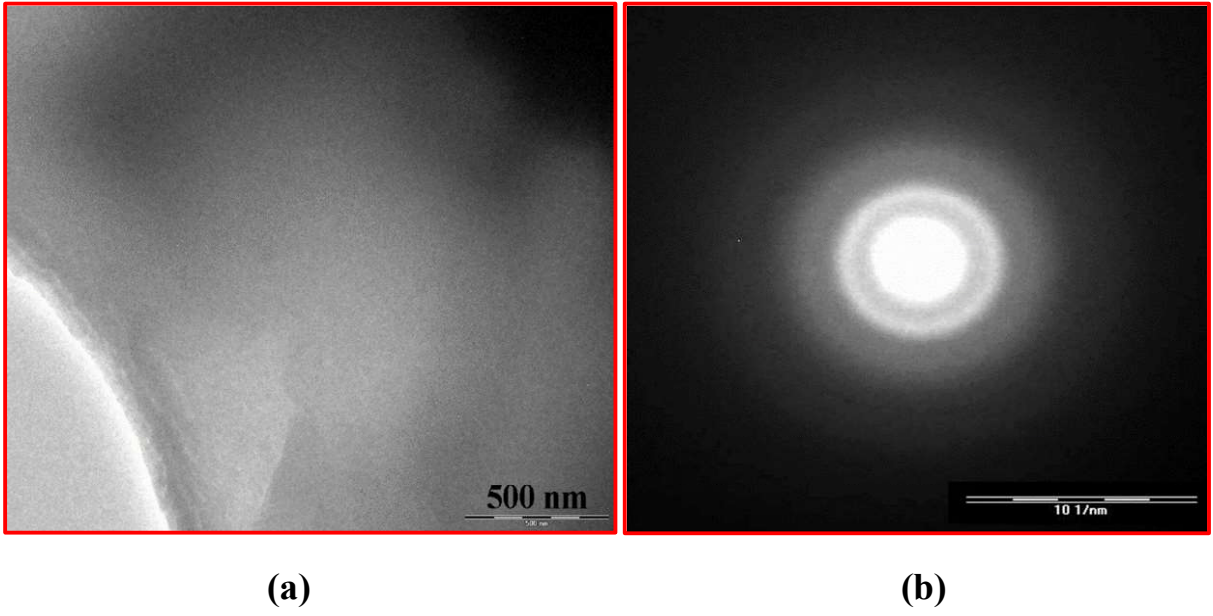

**(a)** **(b)**

Figure 2

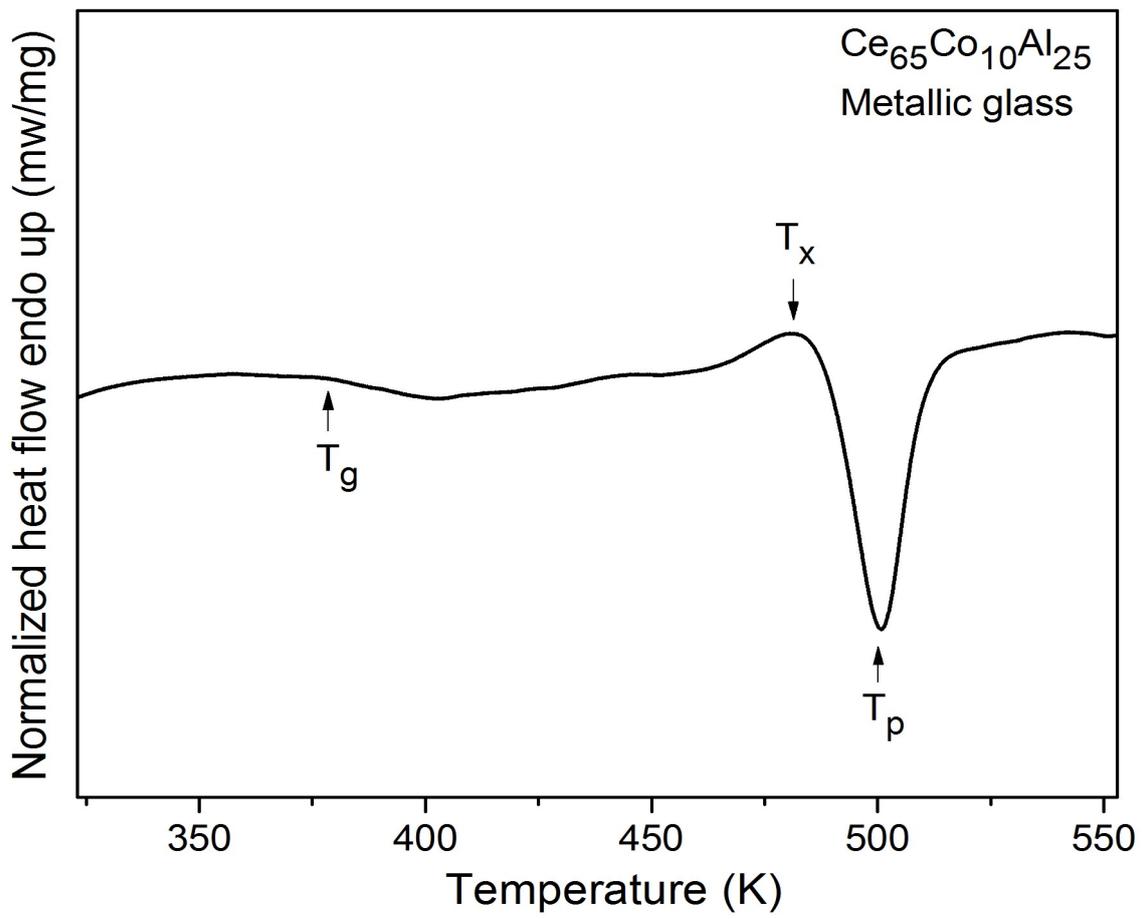

Figure 3

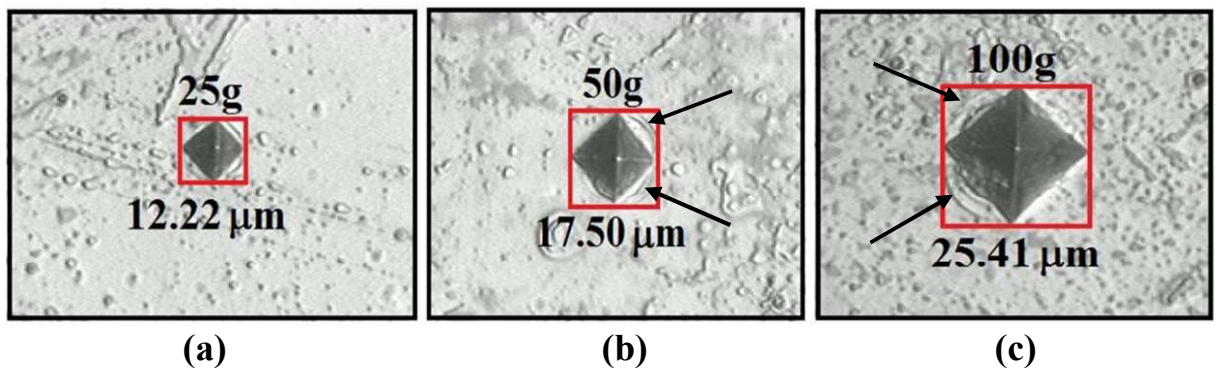

Figure 4

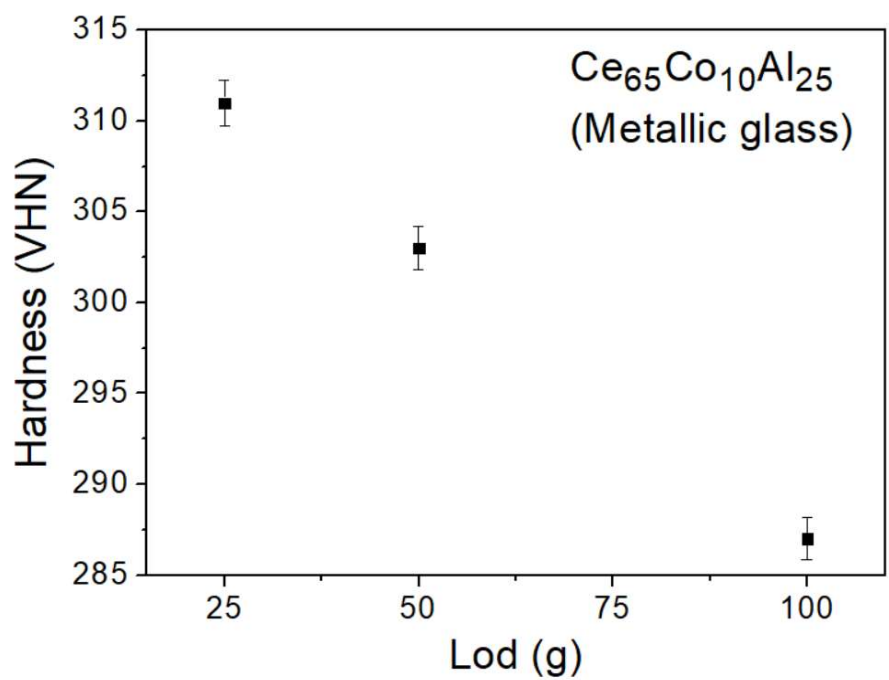

Figure 5

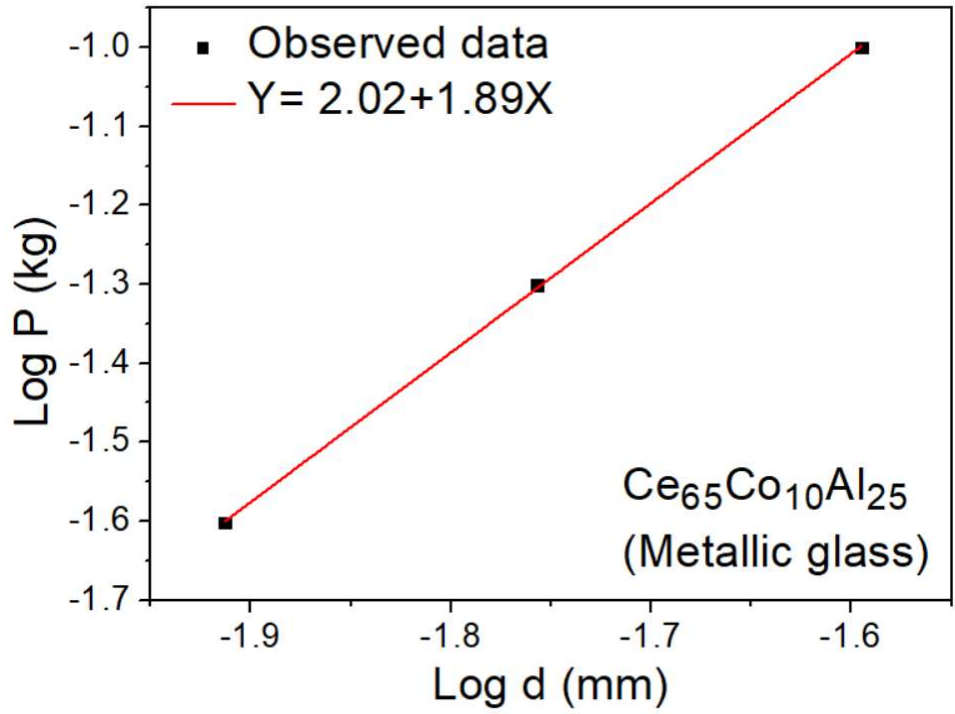

Figure 6